\documentstyle[preprint,prb,aps]{revtex}
\begin{document}

\title{Electrical Noise From Phase Separation In Pr$_{2/3}$Ca$_{1/3}$MnO$_{3}$ Single Crystal}
\author{A. Anane\footnote{Contact author : anane@martech.fsu.edu}, B. Raquet\footnote{Permanent address: LPMC de Toulouse, INSA-SNCMP, Av. De Rangueil. Toulouse, France.} and S. von Moln\'ar} 
\address{MARTECH, Florida State University, Tallahassee, FL 32306-4351}
\author{L. Pinsard-Godart and A. Revcolevschi}
\address{Laboratoire de Chimie des Solides, Université Paris-Sud, 91405 Orsay C\'edex, France}

\maketitle
\tightenlines
\begin{abstract}
Low frequency electrical noise measurements have been used to probe the electronic state of the 
perovskite-type manganese oxide Pr$_{2/3}$Ca$_{1/3}$MnO$_{3}$ versus temperature and in the 
vicinity of the field-induced transition from the insulating, charge-ordered state (I-CO) to the
 metallic, ferromagnetic state (M-F). At high temperature we have observed a high level of the 
excess noise with mainly a gaussian distribution of the resistance fluctuations, and the 
associated power spectral density has a standard 1/$f$ dependence. However, in the hysteretic 
region, where the electrical resistance depends dramatically on the sample history, we have 
observed a huge non-gaussian noise characterized by two level fluctuator-like switching (TLS) 
in the time domain. We discuss the origin of the noise in terms of percolative behavior of the 
conductivity. We speculate that the dominant fluctuators are manganese clusters switching between 
the M-F and the I-CO phases.
\end{abstract}

\narrowtext
\clearpage

{\bf INTRODUCTION}\par

The discovery of the so-called colossal magnetoresistance (CMR) effect in manganese oxides and 
the large experimental effort to develop spin-based electronics ("spintronics") have generated 
a revival of interest in these materials$^1$. Whereas LaMnO$_3$ is in insulating antiferromagnet, 
the substituted compound La$_{1-x}$M$_{x}$MnO$_{3}$ (0.2$\leq$x$\leq$0.35) and M is an alkaline 
earth) is a ferromagnet and undergoes a transition from an insulating to a metallic state when it
 is cooled through its magnetic ordering temperature (T$_C$). But at some rational fractional 
values of x such as 1/8, 1/2 and 3/4, the system tends to achieve a spatial ordering of the
 Mn$^{3+}$, Mn$^{4+}$ ions. This insulating charge-ordered state (I-CO) is often antiferromagnetic.
 However, the CO phase can occur for an extended range of x if the ferromagnetic double-exchange 
interaction is weakened enough by using a rare earth and an alkaline earth of relatively small 
ionic radius. That allows us to establish a desired equilibrium between the different energy 
scales in the system and drive the compound through different ground states via adequate thermal 
and field cycling. A striking example is the system presented in this report, 
Pr$_{1-x}$Ca$_{x}$MnO$_{3}$ (x = 0.3) which is known$^{2,3}$ to undergo a CO phase transition 
bellow T$_{CO}\sim$240K. The magnetic ground state is a canted ordering of Mn spins. At low 
temperature, the application of a magnetic field induces a transition from insulating to metallic
 state resulting from the forced alignment of the Mn spins. This phase transition is of first 
order and the corresponding (T,H) phase diagram displays a strong hysteresis (Fig. 1). We report 
here the first noise measurement on a charge-ordered manganite.\par
A single-crystal was grown using a floating zone method with an image furnace$^4$. A strip of 
approximately 8$\times$2$\times$1 mm$^3$ was cut. The electrical noise was measured in a variable 
temperature cryostat (1.6 - 350 K) equipped with a 6 Tesla superconducting magnet, with either a 
standard five probe AC technique5 using a PAR 124A lock-in amplifier at a frequency of 480 Hz, or 
a four probe DC technique$^5$ using a SR 560 amplifier, and a HP 35660A spectrum-analyzer. The 
noise power spectral density $(S_V/V^2)$ presented below is corrected for the trivial dependence
 of the power spectral density (S$_V$) on the sample resistance $R$ and the applied current 
$I$ $(S_V \propto I^2R^2)$. Furthermore, we have checked the I$^2$ dependence of S$_V$ confirming
 that the measured noise arises only from the fluctuation of the sample resistance. For all the 
measurements presented here, The I-V characteristics show no deviation from Ohmic behavior.\par

{\bf RESULTS AND DISCUSSION}\par

Figure 2 shows the temperature dependence of the normalized power spectral density $S_V/V^2$. 
Over the entire temperature range presented, the power spectral density has a 1/$f^{ \alpha}$ 
dependence with $\alpha$ close to one. The distribution of the voltage fluctuations is mainly 
gaussian; however, we often observe that for short periods of time (a few seconds) the voltage 
spontaneously undergoes slight non-gaussian fluctuations. A two order of magnitude decrease of 
the noise level is observed when the sample is heated from 120K to T$_{CO}$, followed by a slight
 increase above T$_{CO}$. To compare the noise level in different materials, it is customary to 
use the empirical Hooge formula$^6$
 $\displaystyle S_V/V^2=\frac{\gamma}{\Omega \: n}\cdot\frac{1}{f^{ \alpha}}$, where $f$ 
 is the frequency, $n$ is the density of carriers, $\Omega$ is the noisy volume and $\gamma$ 
is the dimensionless Hooge constant. The value of $\gamma$ in normal metals is in the range 
of 10$^{-3}$-10$^{-2}$ [cf. Ref. 7]. In a recent study$^8$, the value of $\gamma$ in 
La$_{2/3}$Sr$_{1/3}$MnO$_{3}$ system has been estimated to be 10$^2$. Because of the 
1/$\Omega$ dependence of the noise level, it is usually impossible to measure the 1/$f$ noise 
in bulk samples. Unfortunately, the effective density of carrier in
 Pr$_{2/3}$Ca$_{1/3}$MnO$_{3}$ is unknown. However if we assume a density of carriers of 1/3 per
 formula unit (corresponding to the doping level), $\gamma$ would be of the order of 5$\times
 $10$^6$ at 300 K. Such a high value does not fit into the general understanding of the excess
 noise. Actually, the 1/$f$ noise is usually attributed to a superposition of a large number of 
fluctuators modeled by a two level process (TLP) with a broad distribution of activation energies. For a given fluctuator the coupling to the resistivity is different in each well of the TLP. In metals, the noise is attributed to thermally activated motion of structural defects$^{9,10}$. In oxides, including the manganites, the oxygen vacancies may be the origin of the large level of noise in those systems$^{11}$. But to explain such a high noise level in Pr$_{2/3}$Ca$_{1/3}$MnO$_{3}$ we would need to invoke an anomalously high density of defects and/or a very strong coupling of each defect to the resistivity. To our knowledge the highest value of $\gamma$ ever reported was 10$^8$ in a In$_2$O$_{3-x}$ film close to the electrical percolation threshold$^{12}$. Indeed, the noise theory in metal-insulator transitions predicts a divergence of the power spectral density when the connectivity of the network is close to the percolation threshold$^7$. Clearly, we are not in the classical scenario of a metal-insulator transition, but a reasonable explanation would be to ascribe the noise in Pr$_{2/3}$Ca$_{1/3}$MnO$_{3}$ to non-homogeneous current transport. In this picture, most of the noise arises from the narrowest current paths; just a few of those critical bonds in a percolative network can influence the connectivity of the entire lattice and therefore induce a large level of noise. When the temperature is decreased, the I-CO phase is stabilized, leading to a decrease of the percolative paths and therefore to an increase of the resistivity. In this situation, the noise level will increase because the net influence on the resistance fluctuations of the remaining current paths will increase. Ultimately, one could expect that for low enough temperatures, we may observe a crossover from a gaussian behavior to TLS of the resistance fluctuations because we will be able to resolve individual TLP. Unfortunately, the resistivity of the sample is too large to make noise measurements below 100 K.\par
To resolve noise behavior when the sample is in a mixed phase configuration (Fig. 1) we have 
performed the following experiment: The sample was first slowly cooled from room temperature 
in zero field down to liquid helium temperature. Then the field was applied. At low field the 
sample was an insulator and its resistance exceeded the measuring range (300 M$\Omega$) of the 
Ohm-meter. At a well-defined field corresponding to the line boundary between the I-CO phase and 
the M-F phase the resistance dropped by several order of magnitudes. The field was then quickly 
set to zero.\par
The effect of the field is to "melt" the charge-ordered state. Recently, a time-aftereffect study
 on the resistance and on the magnetization on the same single crystal has established that there
 is a spatial distribution of transition fields$^{13}$, and the field-induced metal insulator
 transition arises when a percolative network of conductive clusters is formed between the 
contacts. The fact that in the present study the field was cut off just after the sample started 
to conduct the current, leads us to think that the sample is in the vicinity of the percolation 
threshold.\par
When performing noise measurement on non-homogeneous systems, special care has to be taken. The 
contact resistance can be either much lower or much higher than the probed resistance. If the
 first case is not of a concern, having high contact resistance can strongly affect noise 
measurement. The effect on the current noise due to the fluctuation in the contact resistance 
is strongly reduced by using a large bias resistance. But we also encountered the situation 
where the two-point resistance between the current contacts was of the order of a few kilo-Ohms 
while the resistance between the voltage contacts was above our measuring range. This indicates 
that the current does not cross the contact areas of the voltage probes. Fortunately, the current 
path seems to be different after each thermal cycling. This implies that the inhomogeneities are 
not related to physical defects or chemical disorder in the sample, but rather to a statistical 
probability to be in a given state out of many possible configurations with comparable 
energies.\par
Figure 3.a shows the resistance fluctuations in the time domain after applying the procedure 
described above. Clearly resolved resistance switches are evidenced with a relative variation 
of the resistivity as large as 20\% (the largest ever reported) in this bulk sample. Another 
remarkable feature is the qualitative difference between the two traces taken just a few seconds apart under identical experimental conditions. Trace (a) shows a clear switching type noise between 3 levels whereas the non-gaussian behavior of trace (b) is much smaller in amplitude. The TLS behavior has never been stable long enough to perform statistics on the life-time of the states.\par
A similar behavior of the noise - intermittent on-set of telegraph noise - has been observed in 
Si:H system$^{14}$ and interpreted as a signature of inhomogenious current filaments subject to 
localized diffusion processes$^{15}$. The power spectral density in our case has roughly a 1/$f$ 
dependence, but its amplitude fluctuated from one measurement to another. The 1/$f$ dependence is 
due to the way we measure S$_V$; indeed, the usual way to reduce the noise is to average over a 
large number of spectra recorded serially. As the non-gaussian noise appears only from time to 
time, the frequency dependence of S$_V$ is not significantly changed. But the effect of the 
non-gaussian behavior on the amplitude of S$_V$ is dramatic, because the switching behavior can
 produce relative changes in the resistivity of order 10$^{-1}$ while for the gaussian noise this
 change is more of order 10$^{-5}$.\par
Finally, some comments on the physical origin of the fluctuators responsible for the observed 
noise. First, we believe that the observed noise in the mixed phase state and the noise at higher 
temperature are of the same nature; i.e., that they originate from the same kind of fluctuators. 
Secondly, the large amplitude of the steps in the switching regime leads us to speculate that the
 fluctuators are manganese clusters located in the percolation path switching between M-F phase
 and I-CO phase. Therefore, the large level of noise in Pr$_{2/3}$Ca$_{1/3}$MnO$_{3}$ is due to 
a phase separation. Furthermore, the phase separation does not seem to be related to any chemical
 inhomogeneitie of the sample, which leads us to conclude that the phase separation is of an 
electronic type.\par
In summary, we have performed noise measurement on a charge ordered manganite system 
(Pr$_{2/3}$Ca$_{1/3}$MnO$_{3}$). We have observed a huge increase of the level of noise as the CO 
state is stabilized. We have performed noise measurement in a mixed-phase state and observed 
intermittent switching-like noise with amplitudes comparable to 20\% of the resistance value. We
 have interpreted our results in terms of percolative behavior of current transport due to an 
electronic phase separation.\par

{\bf ACKNOWLEDGMENTS}\par

This work has been supported by DARPA and the office of Naval Research under Contract
 No. ONR-N00014-96-1-0767. The Laboratoire de Chimie des Solides is Unité de Recherche Associ\'ee
 au Centre National de Recherche Scientifique No. 446.
\clearpage
{\bf REFERENCES}\\ \\
$^1$	J. M. D. Coey, M. Viret and S. von Moln\'ar, Adv. Phys. {\bf 48}, 167 (1998). See also,
 A.P. Ramirez, J. Phys.: Condens. Matter {\bf 9}, 8171 (1997).\\
$^2$	Y. Tomioka et al., Phys. Rev. B {\bf 53}, R1689 (1996).\\
$^3$	Z. Jirak et al., J.Magn. Magn. Mater. {\bf 53}, 153 (1985).\\
$^4$	A. Revcolevschi and R. Collongues, C. R. Seances Acad. Sci., Ser. A. {\bf 266}, 1767 (1969).\\
$^5$	J. H. Scofield, Rev. Sci. Instrum. {\bf 58}, 985 (1987).\\
$^6$	F. N. Hooge, Phys. Lett. {\bf 29A}, 139 (1969).\\
$^7$	Sh. Kogan, Electronic noise and fluctuations in solids, Cambridge University Press (1996). \\
$^8$	B. Raquet, J. M. D. Coey, S. Wirth, S. von Molnár. Phys. Rev. B {\bf 59}, 12435 (1999). \\
$^9$	P. Dutta and P. M. Horn, Rev. Mod. Phys. {\bf 53}, 497 (1981).\\
$^{10}$M. B. Weissman, Rev. Mod. Phys. {\bf 60}, 537 (1980).\\
$^{11}$M. Rajeswari et al., Appl. Phys. Lett. {\bf 73}, 2672 (1998).\\
$^{12}$O. Cohen and Z. Ovadyahu, Phys. Rev. B {\bf 50}, 10442 (1994). \\
$^{13}$A. Anane, J.-P. Renard, L. Reversat, C. Dupas, P. Veillet, M. Viret, L. Pinsard and
 A. Revcolevschi, Phys, Rev. B {\bf 59}, 77 (1999).\\
$^{14}$C. E. Parman, N. E. Israeloff and J. Kakalios, Phys. Rev. B {\bf 44}, 8391 (1991).\\
$^{15}$L. M. Lust and J. Kakalios, Phys. Rev. Lett. {\bf 75}, 2192 (1995).\\
\clearpage
{\bf FIGURE CAPTION}\\ \\

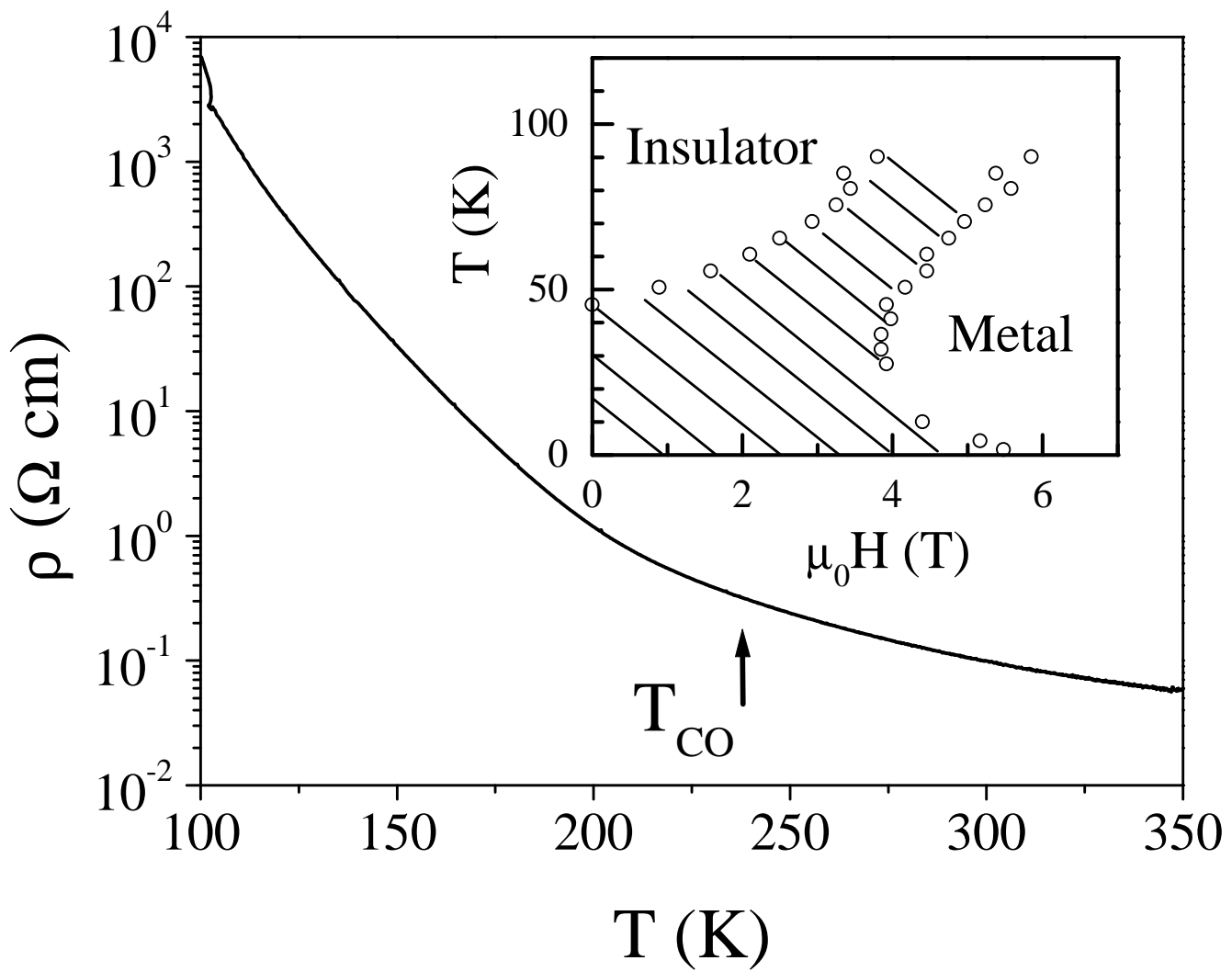
Fig.1 : Resistivity vs. temperature in zero-field. The phase diagram in the (H,T)-plane is 
shown in the insert. Notice the large hysteresis region (the hatched area) in which the I-CO 
and the M-F can coexist.\\ \\

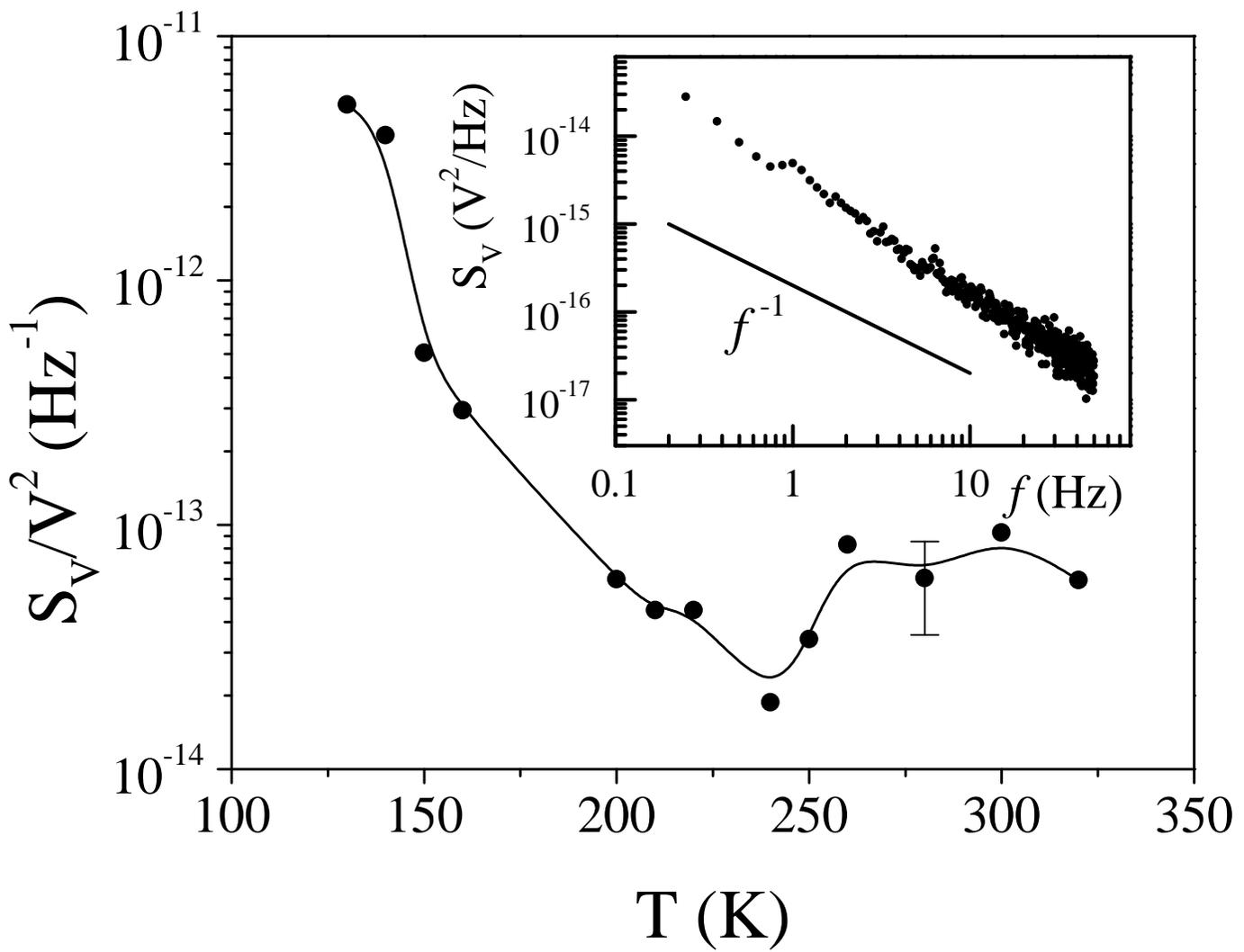
Fig.2 : Temperature dependence of the normalized noise power spectral density in
 Pr$_{2/3}$Ca$_{1/3}$MnO$_{3}$ at 2 Hz. The solid line is a guide to the eye. The insert is the
 frequency dependence of the power spectral density at 300 K showing a 1/$f^{ 1.3}$ dependence.
\\ \\

Fig.3 : Example of the fluctuation voltage as a function of time at 4 K and zero field.
 The applied current was 33 $\mu$A. See text for details.

\end{document}